\documentclass[12pt,a4paper,twoside]{article}
\usepackage{epsfig}
\usepackage{graphics}
\usepackage{color}
\newlength{\figwidth}
\newlength{\figheight}
\def\z0{Z}

\title{
       \vspace{-1.5cm}
       \begin{flushright}
       \begin{tabular}{l}
       {\large CERN-TH/2002-285 }    \\[-3mm]
       {\large IFT - 20/2002}    \\[-3mm]
       {\large hep-ph/0211112}\\
       {\large September 2002}
       \end{tabular}
       \end{flushright}
       \vspace{1.5cm}
      \sc  
Longitudinal virtual photons 
and the interference terms in  $ep$ collisions
}

 \author{Urszula Jezuita-D\c{a}browska $^a$ and Maria Krawczyk $^{a,b}$ \\
 {\small\it a)  Institute of Theoretical Physics, Warsaw University,
    ul. Ho\.za 69, 00-681 Warsaw, Poland} \\[2mm]
{\small\it b) Theory Division, CERN, CH-1211 Geneva 23, Switzerland} \\
 }

\date{}

\begin{document}
		
\maketitle

\vfill

\begin{abstract}
The importance of the contributions of 
the longitudinally polarized virtual photon 
$\gamma^{\ast}_L$
in  $e p$ collisions is  investigated.
We derive the factorization formulae for the unpolarized inclusive and
semi-inclusive $e p$ collisions in an arbitrary reference frame.
The numerical calculations for  the 
large-$p_T$ (prompt) photons production in the 
unpolarized Compton process $e p \rightarrow e \gamma X$ at  the $ep$ HERA  
collider are performed in the Born approximation.
We studied various distributions
in the $ep$ centre-of-mass frame and found that
the differential cross section for the longitudinally 
polarized intermediate photon, $d \sigma_L$,
and the term due to the interference between 
the longitudinal- and transverse-polarization states of the photon,
$d \tau_{LT}$,
are small, i.e. below $10 \%$ of the cross section.
Moreover, these two contributions
almost cancel one  another, leading to a stronger domination of
the transversely polarized  virtual photon, even for 
its  large virtuality $Q^2$. 
Relevance of the resolved longitudinal photon in a 
jet production in DIS events at HERA is commented.
A relatively large ($\sim 30 \%$) effect due to  
 the interference term $d \tau_{LT}$  was found
in the azimuthal-angle distribution in the Breit frame. 
\end{abstract}

\vfill
\begin{flushleft}
CERN-TH/2002-285 \\
IFT - 20/2002    \\
hep-ph/0211112 \\
September 2002
\end{flushleft}
\thispagestyle{empty}

\section{Introduction}
\label{sec:mkj:111}
~\newline
\vspace{-1cm}

Assuming that the one-photon exchange dominates in the deep inelastic 
lepton--nucleon collisions (DIS), a cross section for such a process 
can be described 
in terms of two transverse (T) and one longitudinal (L)
polarization states of the intermediate virtual photon $\gamma^{\ast}$.
The differential 
cross section for the unpolarized $l N \rightarrow l X$ process
can always be decomposed into two differential cross sections, 
$d \sigma_T$ and $d \sigma_L$,
describing the processes with the transversely and
the longitudinally polarized $\gamma^{\ast}$,
$\gamma^{\ast}_T N \rightarrow X$ and $\gamma^{\ast}_L N \rightarrow X$,
respectively \cite{Hand:1963bb}--\cite{Budnev:1975}.
When  the initial particles in  the  discussed process
 $l N \rightarrow l X$
are polarized, 
or when for the unpolarized particles  a semi-inclusive process
$l N \rightarrow l a X$ is considered, there appear  in addition terms 
coming from the interference between the longitudinally and transversely 
polarized virtual photons, 
and  between two different transverse-polarization states of 
$\gamma^{\ast}_T$,  $d \tau_{LT}$ and  $d \tau_{TT}$, respectively
  \cite{Budnev:1975}.

It is well known that  for
the two-photon exchange processes, for instance in the 
$e^+ e^-$ collisions, the interference terms
occur in the cross sections, 
as discussed in \cite{Budnev:1975, Arteaga:1995px}\footnote{
In  $e^+ e^-$ collisions, the interference terms
are also important for the Higgs boson production via $WW$ or $ZZ$
fusion as  was shown in \cite{Dobrovolskaya:1991kx, Bambade:1993vw}.}.
The detailed study of relevance of various contributions, especially
of the interference terms, has been performed in 
\cite{Abbiendi:1999pv, Nisius:2000cv} for
the process $e^+ e^- \rightarrow e^+ e^- \mu^+ \mu^-$,
for the kinematical range of the PLUTO and LEP experiments.
For a corresponding subprocess
$\gamma^{\ast} \gamma^{\ast} \rightarrow \mu^+ \mu^-$
large cross sections for the contribution 
involving at least one longitudinally polarized photon were found.
Moreover, the interference terms 
were found to give  a large  negative contribution.
Both contributions vary strongly as a function of the kinematical variables,
and for some kinematical regions
a cancellation between
the  cross sections for processes
with one or two $\gamma_L^{\ast}$ 
and the interference contributions occurs.
The conclusion from this analysis was that  
 both  types of contributions
have to be taken into account in  extracting from the data
 the leptonic (mionic) structure functions of the virtual photon
 (see also \cite{Krawczyk:2000nh}).
However, in some of the  measurements of the structure functions
$F_2^{\gamma^{\ast}}$ and $F_L^{\gamma^{\ast}}$,
the interference terms
and cross section for two longitudinally 
polarized virtual photons  were neglected, see 
\cite{Berger:1984xu, L3:2000}, and \cite{Schienbein:2002wj}.\\

The contributions due to the longitudinally polarized virtual photons
occur also in  electroproduction and 
the question is how  significant  such   contributions are.
Moreover, this  question is related to the   ongoing discussion on
the relevance of the 
resolved-$\gamma^{\ast}_L$ contributions in the hard
processes \cite{Friberg:2000zz}--\cite{Chyla:2000cu}.
For example, it was pointed out in \cite{Chyla:2000cu} that,
in the case of the dijet production in the $ep$ HERA collider,
the  contributions 
coming from the longitudinally polarized photon 
are sizeable and the
partonic content of the $\gamma^{\ast}_L$ should be
taken into account in describing the data.
However, it is clear that  the study 
of cross sections with  $\gamma^{\ast}_L$ 
for the large-$p_T$ processes should be accompanied 
by a consideration of the corresponding interference terms
containing $\gamma^{\ast}_L$. Unfortunately, there is a lack of 
such studies in the literature. In this paper 
we would like to initiate a discussion on the  relevance of such terms 
in the semi-inclusive $ep$ processes.

It is well known that to get access to the
interference between $\gamma^{\ast}_L$ and $\gamma^{\ast}_T$,
or between two different transverse-polarization states of $\gamma^{\ast}$,
it is natural to consider the azimuthal-angle dependence
in a  special reference frame called
the Breit frame
\cite{Brown:1971}--\cite{Ahmed:1999ix}.
Recently, the  azimuthal asymmetries for the charged-hadrons production
in the  neutral-current deep inelastic $e^+ p$
scattering have been measured with the ZEUS detector at HERA 
\cite{Breitweg:2000qh}, 
and indeed effects due to the corresponding interference terms were observed.

In this article we study the contributions
due to $\gamma^{\ast}_L$ and $\gamma^{\ast}_T$ in  unpolarized $ep$
collisions at  HERA,  
taking as an example the process 
with a production of  high-$p_T$ (prompt) 
photon: $e p \rightarrow e \gamma X$ (Compton process).
In Section~\ref{sec:mkj:222} 
a short derivation of the factorization formulae for
the inclusive and semi-inclusive $ep$ processes is presented
in an arbitrary frame (some details are given also in  Appendix A).
In the case of the semi-inclusive collision,  two
cross sections, $d\sigma_T$ and  $d\sigma_L$, and two
additional terms due to the interference between different polarization 
states of $\gamma^{\ast}$, $d\tau_{LT}$ and $d\tau_{TT}$, appear.
The relation of these interference terms to the contributions proportional to
$\cos \phi$ and $\cos 2 \phi$ in the azimuthal-angle distribution for 
a final $\gamma$ 
in the Breit frame is discussed
in Section~\ref{sec:mkj:333} and Appendix B.
Section~\ref{sec:mkj:444}  
is devoted to the numerical studies of the different contributions
for the  process $e p \rightarrow e \gamma X$
in the Born approximation.
Conclusions are presented in Section~\ref{sec:mkj:555}.
In the Appendices the explicit form of the polarization vectors of
the $\gamma^{\ast}$ and factorization formulae for the semi-inclusive process
are presented both in a frame-independent form and for 
the Breit frame.

\section{Factorization in
             the inclusive and 
semi-inclusive\protect{\\} unpolarized 
            lepton--nucleon scattering}
\label{sec:mkj:222}
~\newline

\vspace{-1cm}
\subsection{Inclusive process $e p \rightarrow e X$ (DIS)}
\label{sec:mkj:222:aaa}
~\newline

\begin{figure}[h]
\begin{center}
\epsfig{file=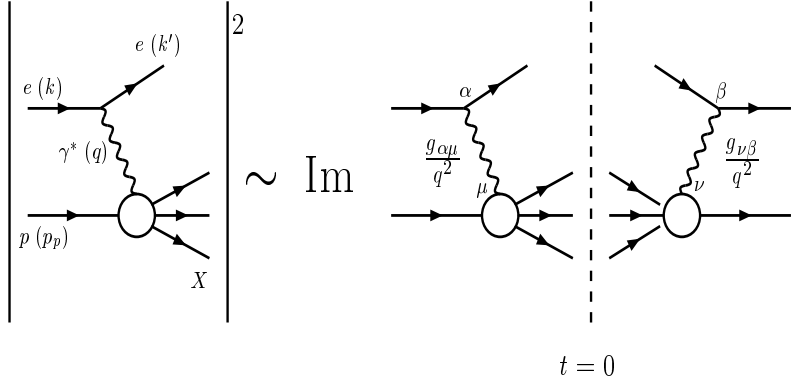, width=18cm,height=6cm}
\end{center}
\caption{Kinematics and notation for the  process 
$e p \rightarrow  e X$ with the one-photon exchange.
The optical theorem relation of the squared matrix element for 
$e p \rightarrow  e X$ and the imaginary part of the amplitude for
$e p \rightarrow  e p$ at $t = 0$.}
\vspace{0cm}                              
\label{fig:mkj:1}
\end{figure}

\vspace{-1cm}
We start with a short description of
the standard DIS process for   unpolarized $e p$ collisions 
(Fig.~\ref{fig:mkj:1}),
\begin{equation}
e p \rightarrow e X \; ,
\label{eq:mkj:p1}
\end{equation}
assuming that the one-photon exchange dominates.
The corresponding differential cross section is denoted by 
$d\sigma^{e p \rightarrow e X}$, and
we use the following notation for the kinematical variables:
$k^{\mu}$ ($k^{\prime \; \mu}$) denotes the four-momentum of the initial 
(final) electron,
$p_p^{\mu}$ the four-momentum of the initial proton,
$q^{\mu}$ the four-momentum of the intermediate photon,
and $Q^2 = -q^2 > 0$ is the photon's virtuality. We denote by 
$\varepsilon^{\mu}_i$ ($i = T_1, T_2, L$) 
the two transverse- and one  longitudinal-polarization vectors of the 
exchanged virtual photon $\gamma^{\ast}$.
The standard scaling variables are $x=Q^2/2p_pq$ and
$y \; = \; p_pq/p_pk$.

~\newline

It is well known that 
for the considered process (\ref{eq:mkj:p1})
there is  a factorization of the differential cross section
onto the lepton and hadron parts and a separation between the 
contributions of the longitudinal-
and transverse-polarization states of the intermediate photon 
 \cite{Hand:1963bb}--\cite{Budnev:1975}. So, we have here:
\begin{equation}
d\sigma^{e p \rightarrow e X} \; = \; 
\Gamma_T \; d\sigma^{\gamma^{\ast} p \rightarrow X}_T \; + \; 
\Gamma_L \; d\sigma^{\gamma^{\ast} p \rightarrow X}_L \; \equiv \;
d\sigma_T \; + \; d\sigma_L \;,
\label{eq:mkj:t1}
\end{equation}
where $d\sigma^{\gamma^{\ast} p \rightarrow X}_T$ and 
$d\sigma^{\gamma^{\ast} p \rightarrow X}_L$ are the cross sections for the
$\gamma^{\ast} p$ collision with the virtual photon polarized transversely 
and longitudinally, respectively. The functions
$\Gamma_T$ and $\Gamma_L$ describe
 the probabilities of the emission, by the initial electron,
of a virtual photon in the transverse- and the longitudinal-polarization 
states, respectively.\\

The above factorization and separation formula 
can be obtained in  various  ways
\cite{Hand:1963bb}--\cite{Budnev:1975}.
For example, the cross section for the process $e p \rightarrow e X$
(see Fig.~\ref{fig:mkj:1})
can be expressed as a  convolution of 
the lowest order leptonic tensor $L_e^{\mu\nu}(k, q)$ and 
the hadronic tensor $W_p^{\mu\nu}(p_p, q)$, 
both symmetric in the indices $\mu$ and $\nu$.  Namely we have 
(for $k^2 = k^{\prime \; 2} = 0$, $p_p^2 = M^2$)
\begin{equation}
d \sigma^{e p \rightarrow e X} \; \sim \; \frac{1}{q^4} \;
L_{e \; \mu\nu} \; W_p^{\mu\nu} \; ,
\end{equation}
where:
\begin{equation}
L_{e}^{\mu\nu} \; (k, q) \;\; = \;\; 2 \; (2 k^{\mu}k^{\nu} - q^{\mu}k^{\nu} - k^{\mu}q^{\nu} + \frac{1}{2}q^2g^{\mu\nu} ) \; ,
\label{ll}
\end{equation}
\begin{equation}
W_{p}^{\mu\nu} \; (p_p, q) \;\; = \;\; W_1 \; (-g^{\mu\nu} + \frac{q^{\mu}q^{\nu}}{q^2}) \; + \; \frac{W_2}{M^2} \; (p_p^{\mu} - \frac{p_pq}{q^2} q^{\mu})(p_p^{\nu} - \frac{p_pq}{q^2} q^{\nu}) \; .
\end{equation}
The  gauge invariance leads to the conditions:
\begin{equation}
q_{\mu} L_{e}^{\mu\nu} \; = \; q_{\nu} L_{e}^{\mu\nu} \; = \; 0 \; , \;\;\;
q_{\mu} W_{p}^{\mu\nu} \; = \; q_{\nu} W_{p}^{\mu\nu} \; = \; 0 \; .
\end{equation}
On the other hand, one can  express the hadronic tensor in terms
of the polarization states of the exchanged photon.
Using the explicit form of the longitudinal-polarization vector
and the completeness relation 
given in Appendix A, we obtain

\begin{equation}
W_p^{\mu\nu} \; = \; 
W_1 \; \sum_{T=T1}^{T2} \; \varepsilon_T^{\ast \; \mu} \varepsilon_T^{\nu} \; + \; (\bar{W}_2 - W_1) \;
\varepsilon_L^{\ast \; \mu} \varepsilon_L^{\nu} \; , \;\;\;\;\;\;
\bar{W}_2 \; = \; \frac{(p_pq)^2 - q^2p_p^2}{-q^2} \; \frac{W_2}{M^2} \; .
\label{eq:mkj:WW}
\end{equation}
From the above form of $W_p^{\mu\nu}$ (\ref{eq:mkj:WW}) one can 
easily derive  formula (\ref{eq:mkj:t1}).\\

Another way of obtaining the considered formula 
(\ref{eq:mkj:t1}) 
is ``the propagator decomposition method'' \cite{Kessler:1970ef}, \cite{ujd:1999}.
In this method
the cross section for  process (\ref{eq:mkj:p1}) is represented
in the following form:
\begin{equation}
d \sigma^{e p \rightarrow e X} \; \sim \; L_e^{\alpha\beta} \; \frac{g_{\alpha\mu}}{q^2} \; \frac{g_{\nu\beta}}{q^2} \; W_p^{\mu\nu} \; ,
\label{eq:mkj:t2}
\end{equation}
where $\frac{g_{\alpha\mu}}{q^2} \; \frac{g_{\nu\beta}}{q^2}$ represents
the propagators of the exchanged photon in the Feynman gauge. 
One can decompose two propagators occurring in eq. (\ref{eq:mkj:t2}) by using
the completeness relation (\ref{ap3}). This leads straightforwardly to
the factorization 
of the cross section for the considered process and, after some calculations, to the
separation into two parts related  to $\gamma_L^{\ast}$ and
$\gamma_T^{\ast}$.
This method is especially useful in analyzing  the semi-inclusive 
processes\footnote{In the case of the semi-inclusive processes, one can 
also use the first method, but then the explicit form of the hadronic 
vertex has to be known
(see \cite{Gourdin:1961}).},
which we will discuss below.

\subsection{Semi-inclusive process $e p \rightarrow e \gamma X$
(Compton process)}
\label{sec:mkj:222:bbb}
~\newline

\vspace{-1cm}
Let us now consider the semi-inclusive process $e p \rightarrow e a X$, 
assuming that all particles are  unpolarized.
Here, 
in comparison to  the DIS process $e p \rightarrow e X$,
one additional particle $a$ is produced.
In the following 
we choose as  particular final state  a 
prompt photon (i.e. $a = \gamma$), with  four-momentum $p$.

\begin{figure}[h]
\begin{center}
\epsfig{file=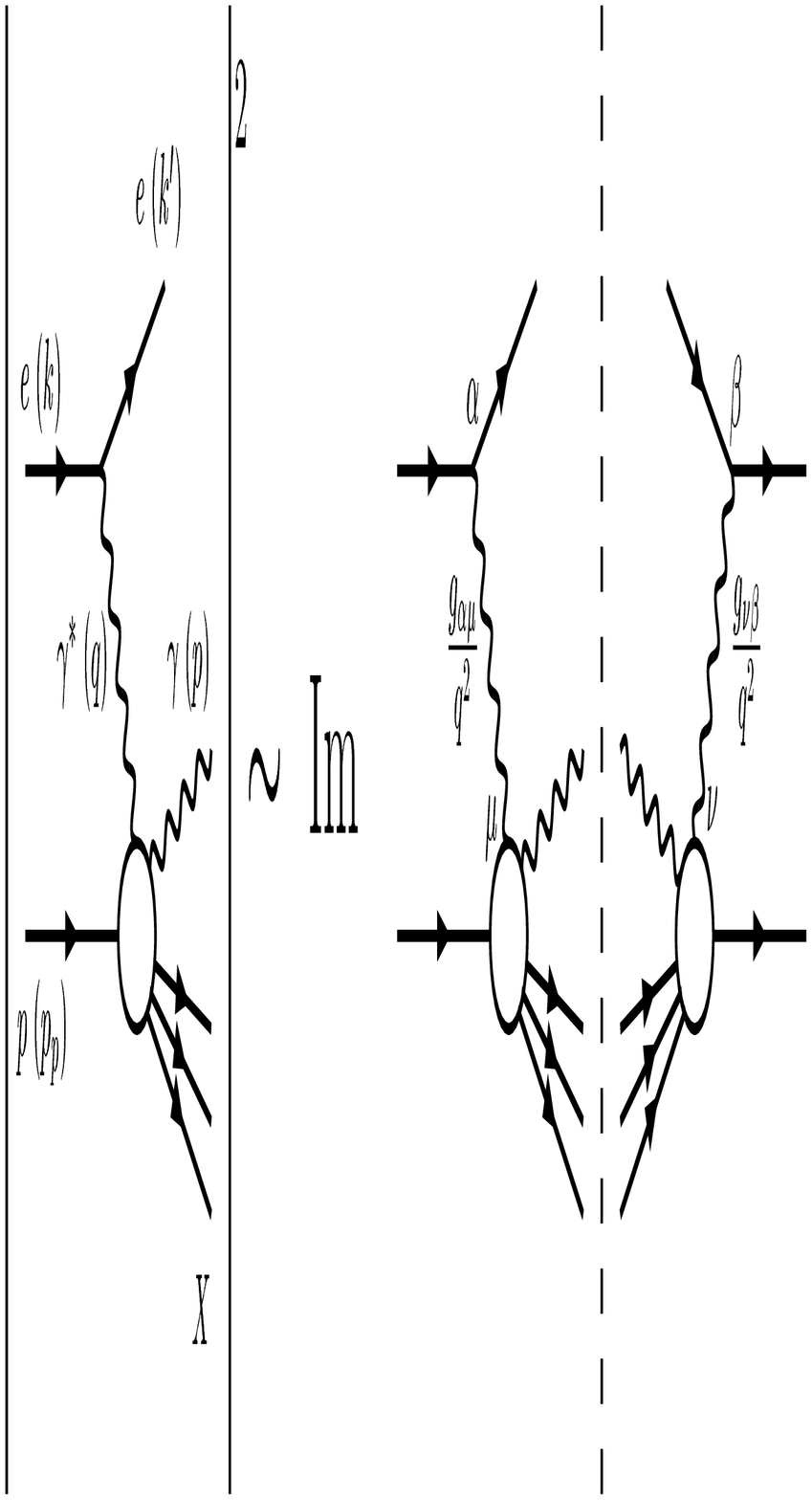, width=18cm,height=6cm}
\end{center}
\caption{Kinematics and notation for the  semi-inclusive process 
$e p \; \rightarrow \; e \gamma X$
with a one-photon exchange.
The optical theorem which relates
$\sigma(e p \rightarrow e \gamma X)$ to the imaginary part 
of the forward amplitude for the process 
$e p \stackrel{\gamma}{\longrightarrow} e p$ $(t = 0)$ 
is also shown.}
\vspace{0cm}                              
\label{fig:mkj:2}
\end{figure}

We will study the factorization of the cross section
for this process, limiting ourselves to the case in which
the $\gamma$ is
emitted from the hadronic part of the diagram only (Fig.~\ref{fig:mkj:2}).
Of course, the final photon  can be emitted
also from the electron line -- a typical  bremsstrahlung process
also  called  the  Bethe--Heitler process;
a relevance of this contribution is discussed at  the 
beginning of Section~\ref{sec:mkj:444}.\\
 
The differential cross section for the unpolarized process
\begin{equation}
e p \rightarrow e \gamma X  \; 
\label{eq:mkj:p2}
\end{equation}
can be written as for  process  (\ref{eq:mkj:p1}), namely
\begin{equation}
d \sigma^{e p \rightarrow e \gamma X} \; \sim \; L_e^{\alpha\beta} \; 
\frac{g_{\alpha\mu}}{q^2} \; \frac{g_{\nu\beta}}{q^2} \; T^{\mu\nu} \; ,
\end{equation}
where the corresponding hadronic tensor $T^{\mu\nu}(p_p, q, p)$
is introduced (cf. (\ref{eq:mkj:t2})).
Here the hadronic tensor $T^{\mu\nu}(p_p, q, p)$
depends not only on the four-momenta of the
intermediate photon $q$ and of the proton $p_p$,
but also on the four-momentum of the final photon $p$.
New scaling variables appear here, e.g.  $z_{\gamma} \; = \; p_pp/p_pq$.
Using ``the propagator decomposition method'' one can obtain
the factorization formula in which
the interference between two 
different transverse-, and 
between the transverse- and the longitudinal-polarization
states of the exchanged photon, denoted by TT  and LT (or TL),  
may appear. We obtain:
\begin{equation}
d \sigma^{e p \rightarrow e \gamma X} \; = \;
\sum_{T=T1}^{T2} \; \Gamma_T \; d\sigma^{\gamma^{\ast} p \rightarrow \gamma X}_T \; + \;  
\sum_{i,j=T1,T2 ; i \neq j} \; \Gamma_{ij} \;d\tau^{\gamma^{\ast} p \rightarrow \gamma X}_{ij} \; + \;
\end{equation}
\begin{eqnarray}
+ \; \Gamma_L \; d\sigma^{\gamma^{\ast} p \rightarrow \gamma X}_L \; + \; 
\sum_{T=T1}^{T2} \; ( \; \Gamma_{TL} \;d\tau^{\gamma^{\ast} p \rightarrow \gamma X}_{TL} \; + \; \Gamma_{LT} \;d\tau^{\gamma^{\ast} p \rightarrow \gamma X}_{LT} \; ) \; .
\label{eq:f}
\end{eqnarray}
Below we will use the following short notation for the groups of contributions 
which appear in (11) and (12)\footnote{Although the symbols 
$\sigma_T$ and $\sigma_L$ have appeared already in (2) for other process (DIS), this should not lead to any confusion, as in the rest of the paper we consider only the semi-inclusive process (9).}:  
\begin{eqnarray}
d \sigma^{e p \rightarrow e \gamma X} \; \equiv \; 
d \sigma_T + d \tau_{TT} \; + \; d \sigma_L \; + \; d \tau_{LT}.
\label{eq:mkj:t5}
\end{eqnarray}
We see that
the cross section (\ref{eq:mkj:t5}) for the considered process (\ref{eq:mkj:p2}) contains 
$d \sigma_T$, $d \sigma_L$ and in addition two interference terms,
$d \tau_{TT}$ and $d \tau_{LT}$
(see also \cite{Budnev:1975}).
These four terms are related by the optical theorem 
(see Fig. \ref{fig:mkj:2}) to the corresponding amplitudes:
\begin{equation}
d \sigma_{T} \; \sim \; 
\frac{1}{q^4} \sum_{T = T1}^{ T2} \; [ \; (\varepsilon_T^{\ast})_{\mu} \; 
L_e^{\mu\nu} \;  (\varepsilon_T)_{\nu} \; ] \cdot 
[ \; (\varepsilon_T)_{\mu} \; T^{\mu\nu} \; 
 (\varepsilon_T^{\ast})_{\nu} \; ] \; ,
\label{T}
\end{equation}

\begin{equation}
d \sigma_{L} \; \sim \; 
\frac{1}{q^4}[ \; (\varepsilon_L^{\ast})_{\mu} \; 
L_e^{\mu\nu} \;  (\varepsilon_L)_{\nu} \; ] \cdot 
[ \; (\varepsilon_L)_{\mu} \; T^{\mu\nu} \; 
 (\varepsilon_L^{\ast})_{\nu} \; ] \; ,
\label{L}
\end{equation}

\begin{equation}
d \tau_{TT} \; \sim \; 
\frac{1}{q^4}\sum_{i,j = T1, T2; \; i \neq j} \; [ \;  (\varepsilon_i^{\ast})_{\mu} \; 
L_e^{\mu\nu} \;  (\varepsilon_j)_{\nu} \; ] \cdot 
[ \; (\varepsilon_i)_{\mu} \; T^{\mu\nu} \; 
 (\varepsilon_j^{\ast})_{\nu} \; ] \; ,
\label{TT}
\end{equation}

\begin{eqnarray*}
d \tau_{LT} \; \sim \; 
\frac{1}{q^4}\{\sum_{T = T1, T2}  \; [ \; (\varepsilon_L^{\ast})_{\mu} \; 
L_e^{\mu\nu} \; (\varepsilon_T)_{\nu} \; ] \cdot [ \; 
(\varepsilon_L)_{\mu} \; T^{\mu\nu} \; 
 (\varepsilon_T^{\ast})_{\nu} \; ] \; +
\end{eqnarray*}
\begin{equation}
+ \; \sum_{T = T_1, T_2}  \; [ \;(\varepsilon_T^{\ast})_{\mu} \; L_e^{\mu\nu} \; (\varepsilon_L)_{\nu} \; ] \cdot [ \;
(\varepsilon_T)_{\mu} \; T^{\mu\nu} \; (\varepsilon_L^{\ast})_{\nu} \; ] \; \}.
\label{LT}
\end{equation} 
~\newline

It is worth noticing that 
the decomposition of the differential cross section 
$d \sigma^{e p \rightarrow e \gamma X}$
into three components: $d \hat{\sigma}_T = d \sigma_{T} +
d \tau_{TT}$, $d \sigma_L$ and $d \tau_{LT}$, does not depend on the choice 
of the reference frame or of the  basis for  the polarization vectors.
Note that 
in the differential cross section $d \sigma^{e p \rightarrow e \gamma X}$
there are two independent terms 
related to the longitudinal-polarization state of the virtual photon:
$d \sigma_L$ and $d \tau_{LT}$.\\

Obviously the above factorization formula (\ref{eq:mkj:t5}) holds for 
the semi-inclusive process with 
arbitrary final-state  particle $a$.

\section{Azimuthal-angle distribution for 
$e p \rightarrow e \gamma X$}
\label{sec:mkj:333}
~\newline

\vspace{-1cm}
In studies of the process  $e p \rightarrow e \gamma X$
it is useful to consider
the azimuthal-angle ($\phi$) distribution. This angle is defined as
the difference between  the azimuthal angle of the final
electron ($\phi_e$) and that of the final photon $\gamma$ ($\phi_{\gamma}$):
\begin{equation}
\phi \; = \; \phi_e \; - \; \phi_{\gamma} \; .
\label{angle}
\end{equation}

\begin{figure}[h]
\vspace{-14cm}
\hspace{2cm}
\epsfig{file=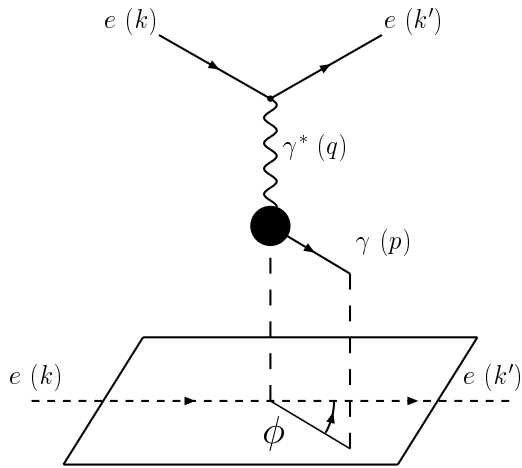}
\vspace{-9cm}
\caption{The azimuthal angle $\phi$ defined in the Breit frame
for the Compton process 
$e p \rightarrow  e \gamma X$.}
\vspace{0cm}                              
\label{fig:mkj:aa}
\end{figure}

In  the special reference frames
in which the momenta of
the virtual photon and of the proton are antiparallel
(for example in the Breit frame or in
the $\gamma^{\ast} p$ centre-of-mass frame)
$\phi$ is the angle between the electron scattering-plane and 
the plane defined by the momenta of the exchanged $\gamma^{\ast}$
and the final  $\gamma$ (Fig.~\ref{fig:mkj:aa}).
In such  reference frames,  the cross section
$d \sigma^{e p \rightarrow e \gamma X} / d \phi$ is linear in 
$\cos \phi$, $\cos 2\phi$, $\sin \phi$ and $\sin 2\phi$.
In the Born approximation
the terms containing $\sin \phi$ and $\sin 2\phi$  vanish
as a  consequence of a time-reversal invariance,
so the azimuthal-angle  distribution  reduces 
to the following form
\cite{Brown:1971}--\cite{Ahmed:1999ix}
(see also Appendix B):
\begin{equation}
\frac{d \sigma^{e p \rightarrow e \gamma X}}{d \phi} \; = \;
\sigma_0 \; + \; \sigma_1 \; \cos \phi \; + \; \sigma_2 \; \cos 2\phi \; ,
\label{eq:mkj:t7}
\end{equation}
with independent on $\phi$ coefficients 
$\sigma_0$, $\sigma_1$ and $\sigma_2$. 
These coefficients
calculated for instance in  the Breit frame, are related to the terms
$d \sigma_T / d\phi$, $d \sigma_L / d\phi$, 
$d \tau_{LT} / d\phi$ and  $d \tau_{TT} / d\phi$, 
introduced in Sect.~\ref{sec:mkj:222:bbb},
calculated in 
 the same reference frame.
In the circular-polarization
basis, where  the polarization vectors correspond to  the definite 
helicity states,
the term $\sigma_2 \cos 2\phi$ arises from the interference between 
two different transverse-polarization states of the  exchanged photon 
($\sim d \tau_{TT}$).
The longitudinal-transverse interference ($\sim d \tau_{LT}$) 
gives rise to the term $\sigma_1  \cos \phi$.
The $\sigma_0$ consists of the sum of
the cross sections with the intermediate  photon polarized transversely
and longitudinally ($\sim (d \sigma_{L} + d \sigma_{T})$).
Obviously, since $\sigma_1$ and $\sigma_2$ (and $\sigma_0$)
do not depend on $\phi$,  
the contributions due to the interferences 
will disappear after integrating over $\phi$.

The interference terms  can be   extracted   in a straightforward way
from the data for the azimuthal-angle distribution $d \sigma / d \phi$
in the frames with antiparallel virtual photon's and proton's momenta,
for example in the Breit frame.

In other frames, i.e. in frames in which the momenta of the virtual photon 
and proton are not antiparallel,
the dependence on the azimuthal angle is more complicated.
Each of the four terms contributing to the differential cross section
(see eqs. (\ref{eq:mkj:t5})--(\ref{LT})) consists of the part that 
does not depend on the azimuthal angle. Consequently the interference terms
integrated over a $\phi$ give non-zero result.
This important
fact was mentioned in papers  
\cite{Dobrovolskaya:1991kx, Bambade:1993vw},
where the Higgs boson production via $WW$ or $ZZ$
fusion in  $e^+ e^-$ collisions has been studied.

\section{Numerical results}
\label{sec:mkj:444}
We perform calculations of the cross sections
for the unpolarized Compton process $e p \rightarrow e \gamma X$.
We consider the one-photon exchange only;
this approximation is correct for the virtuality of the intermediate photon
$Q^2 \ll M_Z^2$,
while for  larger values of $Q^2$ the $Z/W$ boson exchange
should be included.
Below we present predictions
for HERA  for  the beam energies equal to:
$E_e = 27.5$ GeV and $E_p =  920$ GeV. 

We analyze, at the Born level,  the emission of the prompt photon
from the hadronic vertex, i.e. we consider 
the Compton subprocess: $\gamma^{\ast} q \rightarrow \gamma q$.
In principle, the Bethe--Heitler (BH) process, where the  prompt photons 
can be emitted from the electron line
and the interference
between the Compton  and  BH amplitudes,
should   also be included in the calculation 
\cite{Brodsky:1972yx, Brodsky:1999sr, Hoyer:2000mb}.
Both latter contributions 
dominate the cross section in the electron--proton centre-of-mass frame
for the rapidity range of the final photon's $Y < 0$,
while for  greater rapidities the Compton process dominates
\cite{Kramer:1998nb}.
We expect that our results should be reliable
for positive values of the Y in the $CM_{ep}$.\\

Results presented below were obtained for two frames: 
the electron--proton centre-of-mass frame ($CM_{ep}$) and 
the Breit frame. 
The transverse momentum $p_T$ of the final photon is defined
as perpendicular to the direction of the $e p$ collision 
in the $CM_{ep}$ frame or to the direction of the $\gamma^{\ast} p$ collision
in the Breit frame.

For the quark density in the proton $q^p(x,\tilde Q^2)$
 we use the CTEQ5L parton parametrization 
\cite{Lai:2000wy} with a fixed number of flavours: $N_f = 4$. 
This parametrization imposes the limit
on a range of  a hard scale $\tilde Q^2$: it  has to be greater than $1$ GeV
and less than $10^4$ GeV.
In our calculations we use as a hard scale $\tilde Q$
the transverse momentum of the final photon $p_T$; 
effects of other choices of the hard scale were studied by us, and are briefly
discussed below.

\begin{figure}[h]
{\hspace{3.5cm}\input{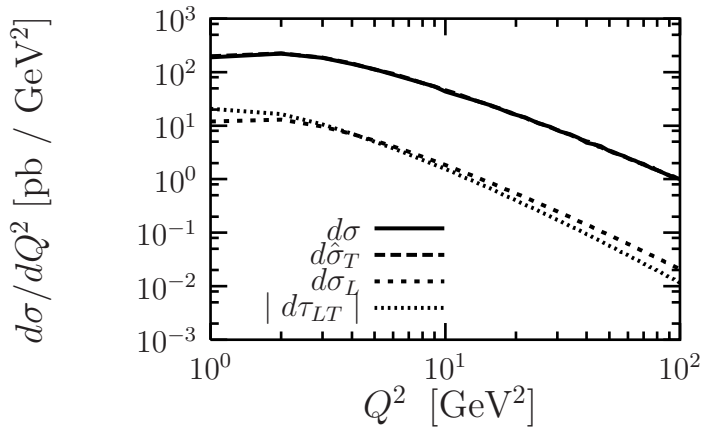}} 
\caption{Various contributions to the differential cross section
$d \sigma / d Q^2$
in the $CM_{ep}$ frame
for the prompt-photon production ($(p_T)_{min}=$~1 ~GeV)
 at HERA.
The contribution due to the transverse-polarization states:
$d \hat{\sigma}_T = d \sigma_T + d \tau_{TT}$ (long-dashed line), 
the cross section for  the longitudinally polarized 
intermediate photon: $d \sigma_L$ (short-dashed line) and
the absolute value of the interference term: $|d \tau_{LT}|$
(dotted line), together with
the sum of all terms $d \sigma$ (solid line), are shown.
}
\label{fig:3}
\end{figure}

First we consider the differential cross section
$d \sigma / d Q^2$
in the $CM_{ep}$ frame, integrated over $p_T$ 
from $(p_{T})_{min} = 1$ GeV  to  $(p_{T})_{max} = \sqrt{S_{ep}}/2$,
as a function of $Q^2$.
We consider separately four  
frame-independent 
contributions:
$d \hat{\sigma}_T = d \sigma_T + d \tau_{TT}$, $d \sigma_L$ and
$|d \tau_{LT}|$ 
as a function of $Q^2$.
As it is presented in   Fig.~\ref{fig:3},
the $Q^2$ dependence is roughly similar for all
contributions.
The differential cross section 
is dominated by the contribution due to the transversely 
polarized intermediate
photon, i.e. $d \hat{\sigma}_T = d \sigma_T + d \tau_{TT}$.
The $d \sigma_L$ contributes at the 
few per cent level to the whole cross section,
similarly to the interference term $d \tau_{LT}$.
Moreover, $d \sigma_{L}$ is positive while  $d \tau_{LT}$  negative,
and these two contributions almost cancel one another.
The resulting cross section $d\sigma/d Q^2$,
even for large values of $Q^2$ ($ \sim 100$ GeV$^2$),
is described 	with a  high accuracy by the
$d \hat{\sigma}_T$ term only. \\

\begin{figure}[h]
\hspace{3.5cm}\input{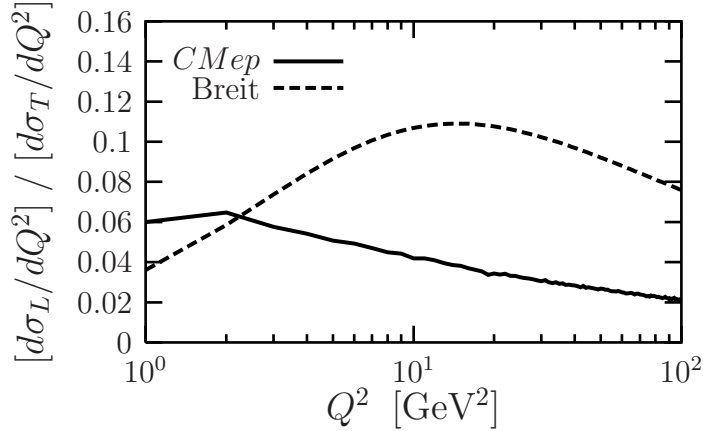} 
\caption{The ratio 
$[d\sigma_L/dQ^2] \; / \; [d\sigma_T/dQ^2]$ 
for prompt-photon production in the $ep$ collision at HERA
($(p_T)_{min}=$ 1 GeV), as a function of $Q^2$, in the $CM_{ep}$ frame (solid line) 
and in the Breit frame (dashed line), is shown.
}
\label{fig:4}
\end{figure}
~\newline

\vspace{-1cm}
The ratio 
$[d\sigma_L/dQ^2] \; / \; [d\sigma_T/dQ^2]$
as a function of virtuality $Q^2$
for the $CM_{ep}$ frame and the Breit frame\footnote{The fixed value
of $p_T$ is changing when we are coming from the $CM_{ep}$ frame to 
the Breit frame.} is presented in  Fig.~\ref{fig:4}.
In the $CM_{ep}$ frame it
grows slowly with the  virtuality of the intermediate photon
up to $Q^2 \sim  (p_{T})_{min}^2$,
then this ratio slowly decreases. 
In the Breit frame 
the corresponding growth is much faster, 
the ratio reaches its maximum at much larger $Q^2$ (10 times $(p_T)_{min}$).
Also the maximum is  higher (two times) 
in the Breit frame than in the   $CM_{ep}$ one.
At  $Q^2 = 100$ GeV$^2$
the considered ratio  $[d\sigma_L/dQ^2] \; / \; [d\sigma_T/dQ^2]$
is equal to about $0.02$ for the $CM_{ep}$ frame, while 
it is about four times larger for the Breit frame.\\

\begin{figure}[h]
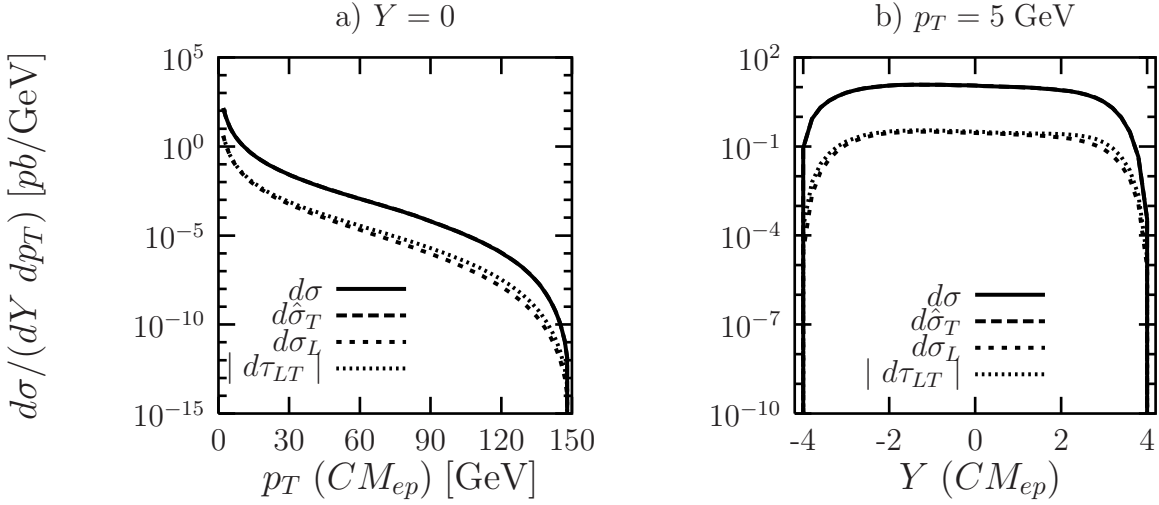

{\input{AepPTY.tex} \input{AepYPT.tex}} 
\caption{Various contributions to the 
$d\sigma / (dY dp_T$) in the $CM_{ep}$  frame
for   high-$p_T$ photon production at
HERA ($(p_T)_{min}=$ 1 GeV) as a function of  a) $p_T$  at $Y=0$
and  b)  $Y$  at $p_T=5$ ~GeV.
Same notation as in Fig.~\ref{fig:3}.
}
\label{fig:5}
\end{figure}
~\newline

\vspace{-1cm}
Figure~\ref{fig:5} shows
results for the differential cross section $d\sigma / dY dp_T$
in the $CM_{ep}$ frame. 
The $p_T$ distribution for a fixed value
of the photon rapidity, $Y=0$, is shown in Fig. \ref{fig:5} (left),  
while  the 
$Y$ distribution at $p_T=5$ GeV is shown in  Fig. \ref{fig:5} (right).
We see that these distributions 
are described with a very good 
accuracy by the $d \hat{\sigma}_T$ terms only. Contributions 
$ d \sigma_L$ and  $d \tau_{LT}$ are small and there is almost 
a cancelation between them. 
In  other words, the measurements of such cross sections
in the $CM_{ep}$ frame
are not sensitive to the individual
contributions
involving the longitudinally polarized virtual photon:
$ d \sigma_L$ and/or $d \tau_{LT}$.

\begin{figure}[h]
\vspace{1cm}
\hspace{3cm}
{\input{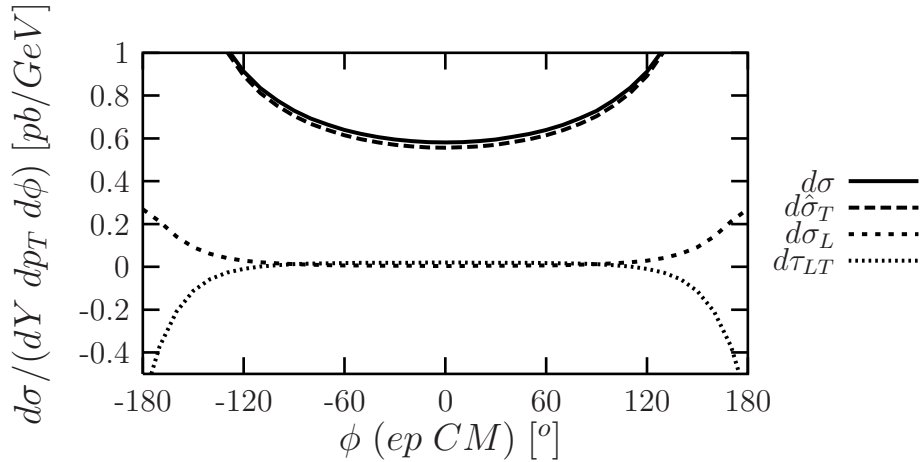}}
\caption{
The $d \hat{\sigma}_T = d \sigma_T + d \tau_{TT}$, $d \sigma_L$ and
$d \tau_{LT}$ contributions to the  high-$p_T$ photon production at
HERA. Results for 
the azimuthal-angle distributions $d \sigma / ( d Y d p_T \; d \phi )$
in the $CM_{ep}$ frame, for
$Y = 0$ and  $p_T = 5$ GeV.}
\label{fig:6}
\end{figure}

Another interesting and important fact is that, 
in the $CM_{ep}$ frame
the interference term
gives non-vanishing contributions even for the  cross sections
integrated over the azimuthal angle (defined by eq.(\ref{angle})),
as it was supposed. This can be seen in 
the azimuthal-angle distribution
for the interference term in this frame (Fig.~\ref{fig:6}).
It is clear that the  $d\tau_{LT}$ integrated over $\phi$ 
gives a non-vanishing contribution.
This is the main difference between 
the $CM_{ep}$   frame and 
the Breit frame (or other   frames with
antiparallel virtual photon and proton).
In the latter frame the contributions due to the interference terms disappear
in the cross sections integrated over $\phi$, see below. \\

Finally, we study the azimuthal-angle distribution of the final photon  
in the Breit frame (19).
As discussed in Sect.~\ref{sec:mkj:333}, in this frame
the coefficients $\sigma_1$, $\sigma_2$ and $\sigma_0$
are independent of $\phi$. They are  directly related to 
the interference terms 
$d \tau_{TT} / d \phi$, $d \tau_{LT} / d \phi$ and to the sum: 
$d \sigma_L / d \phi + d \sigma_T / d \phi$, 
respectively.
(For more details see also  Appendix B).

\begin{figure}[h]
{\hspace{4cm}\input{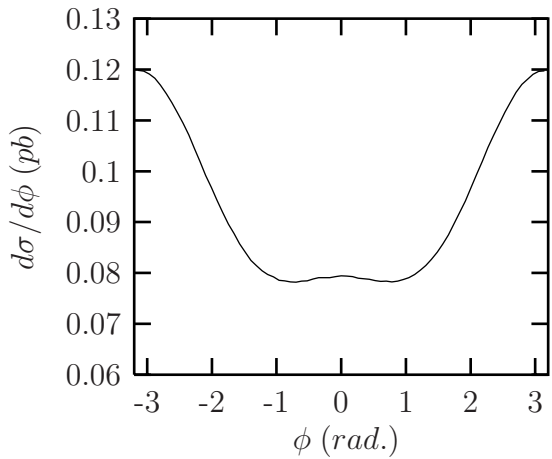}}
\caption{The $\phi$ distributions in 
the Breit frame (see text)
obtained for:
$180$ GeV$^2 < Q^2 < 7220$ GeV$^2$, $0.2 < y < 0.8$, $0.2 < z_{\gamma} < 1.0$
and the $(p_T)_{min} = 2$ GeV, for $e p \rightarrow e \gamma X$ at HERA.
}
\label{fig:mkj:r3}
\end{figure}

The numerical calculations 
for the azimuthal-angle distribution for the Compton process
in the Breit frame
are performed 
for the same kinematical region in which
the charged-hadrons production was measured in the ZEUS experiment
at the HERA collider \cite{Breitweg:2000qh}:
$180$ GeV$^2 < Q^2 < 7220$ GeV$^2$, 
$0.2 < y < 0.8$,
$0.2 < z_{\gamma} < 1.0$
and the $(p_T)_{min} = 2$  GeV.
Results for the Compton process are presented in Fig.~\ref{fig:mkj:r3}.
The contribution related to the interference between two transverse
polarization states of  $\gamma^{\ast}$
(the term proportional to $\sigma_2  \cos 2\phi$)
gives a negligible effect,
while the interference
between the virtual photon polarized longitudinally and transversely
(the term proportional to $\sigma_1 \cos \phi$)
leads to a  visible effect (about $30\%$).
Clearly, both interference contributions being symmetric under 
the $\phi \rightarrow -\phi$ transformation
disappear after the integration over the $\phi$ angle over $-\pi$ to +$\pi$.
\vspace{1cm}

Now we compare our results for the prompt-photon production
with those obtained in the ZEUS experiment for the 
charged hadrons \cite{Breitweg:2000qh}.
In the ZEUS data
the term $\sigma_1 \cos \phi$
is clearly seen for all four considered values of $p_T$ cut:
$(p_T)_{min} \; = \; 0.5$, $1$, $1.5$ and $2$ GeV, while
the term $\sigma_2 \cos 2\phi$ gives a negligible effect
for lower values of  $(p_T)_{min}$,
becoming visible for  a larger   $(p_T)_{min}$.
This is different 
from the Compton process discussed above,
where the corresponding $\sigma_2$ term is very small, 
even for   $(p_T)_{min}=2$ GeV
(see Fig.~\ref{fig:mkj:r3}).
This difference  arises from the following fact:
in the case of the Compton process $e p \rightarrow e \gamma X$,
calculated in the Born approximation,
there is only one subprocess $\gamma^{\ast} q \rightarrow \gamma q$
that  contributes to the cross section.
For the  charged-hadrons production 
there are, at the Born level, two subprocesses, 
$ \gamma^{\ast} q \rightarrow g q$  
and
$ \gamma^{\ast} g \rightarrow q \bar{q}$.
The second process $ \gamma^{\ast} g \rightarrow q \bar{q}$
dominates in the $\sigma_2 \cos 2\phi$ term,
while the contribution coming from the process
$ \gamma^{\ast} q \rightarrow g q$
(analogous to our $\gamma^{\ast} q \rightarrow \gamma q$ process)
dominates in the $\sigma_1 \cos \phi$ term. 
Therefore  both  contributions can (and are) visible in the azimuthal-angle 
distribution of
the  charged-hadrons production, while in our case, based on one subprocess,  
the visible effect is expected only in $\sigma_1 \cos \phi$. \\

Finally, let us comment on the dependence of our results on a choice of 
the hard scale.
Although all the  cross sections  presented above
are computed for
a hard scale equal to  $p_T$, one can choose 
for the considered process 
also other scales: e.g. $Q^2$ or $\sqrt{p_T^2 + Q^2}$.
We have checked by an  explicit calculation that 
the cross sections obtained when the 
hard scale is equal to   $Q^2$ or $\sqrt{p_T^2 + Q^2}$ are
slightly larger (not more than 5\%) 
than the ones obtained when  the hard scale is equal to $p_T$.
This difference is not significant and does not change our main conclusions.

\section{Conclusions}
\label{sec:mkj:555}
~\newline

\vspace{-1cm}
In this paper we investigated the importance of the contributions due to 
the longitudinal virtual photon in  unpolarized, semi-inclusive
$ep$ collisions at HERA. 
In general  there are two such 
contributions: the cross section  
for  $\gamma_L^{\ast}p$ collision $d \sigma_L$,
and the interference term $d \tau_{LT}$.
As a particular semi-inclusive process we chose the Compton process,
where the prompt photon is emitted from the hadronic system. 
For the semi-inclusive  process $e p \rightarrow e \gamma X$
at  HERA energies we analyzed three frame-independent contributions:
the cross section $d \sigma_{L}$,
$d \tau_{LT}$ coming from the interference
between the longitudinal- and the transverse-polarization states of 
$\gamma^{\ast}$ and the contribution due to 
$\gamma^{\ast}_T$:
$d \hat{\sigma}_{T} = d \sigma_T + d \tau_{TT}$. 
The calculations were performed in the Born approximation.
In the $ep$ centre-of-mass frame 
we studied various distributions, the $p_T$ and the rapidity distributions,
as well as the  $Q^2$ distribution,  were the the contributions mentioned 
simply add up.
We found that both $d \sigma_{L}$ and $d \tau_{LT}$
are small and of  similar size, below $10 \%$ of the cross section.
This suggests that 
the contribution $d \sigma_L$ and the interference terms
need  be included on the same footing.
Additionally, because of  their opposite signs,  they almost cancel 
in the cross section.
This leads to  a strong domination of  the considered cross section 
by a contribution due to transversely polarized virtual photon,
even for large values of its virtuality.

Although our results are based on the Compton process in a Born approximation
only, we think that they already shed some light on the
importance of contributions due to $\gamma_L^{\ast}$ in 
other semi-inclusive processes, like  the jet production in the DIS events 
at the HERA collider. 
It is clear that in order to reach a final conclusion  
on the importance of the contributions related to the  $\gamma_L^{\ast}$,
as advocated in some analyses, further studies 
are needed, with the incorporation of the relevant interference 
terms for consistency.

The studies of the azimuthal-angle dependence of
$d \sigma^{e p \rightarrow e \gamma X} / d \phi$
in the Breit frame
give access to the interference 
term $d \tau_{LT}$, as expected.
Its effect is about $30\%$, showing in this case the  importance of the 
$\gamma_L^{\ast}$ contributions in the  form of the  interference 
$d\tau_{TL}$ term. 

\vspace{1cm}
\begin{Large}
\begin{bf}
Acknowledgements
\end{bf}
\end{Large}
~\newline

We would like to acknowledge I.F. Ginzburg, P.M. Zerwas, 
P.J. Mulders, S. Brodsky, L.~Levchuk, and A.~Szczurek
for  interesting and fruitful discussions.\\

This research has been partly supported 
by the Polish Committee for Scientific Research, Grant No 2P03B05119(2000-2002)
and by the European 
Community's Human Potential Program under Contract HPRN-CT-2000-00149
Physics in Collision.

\vspace{1cm}

\noindent
{\huge\bf Appendix}

\label{sec:mkj:666}

\noindent
{\Large\bf Appendix A.}
{\Large\bf Polarization vectors of $\gamma^{\ast}$
            and factorization formula for the Compton process}
~\newline

Below we present the polarization
vectors of a virtual photon $\gamma^{\ast}$ 
(with the four-momentum $q^{\mu}$: $q^2 < 0$)
in a form that  is   independent of the reference frame
for the two  types of basis for  the polarization vectors:
 linear and  circular.

\noindent
{\large\bf A.1 Polarization vectors} 

The Lorentz condition for the polarization vectors gives rise
to the physical observables \cite{}
\begin{equation}
\varepsilon_{\mu} \; q^{\mu} \;\; = \;\; 0 \; .
\label{ap1}
\end{equation}
Consequently, the scalar-polarization state of the $\gamma^{\ast}$,
 described by
the following vector
\begin{equation}
\varepsilon^{\mu}_{S} \;\; = \;\; \frac{q^{\mu}}{\sqrt{-q^2}} \; ,
\label{ap2}
\end{equation}
does not contribute to the physical observables.
We can write the completeness relation as follows:
\begin{equation}
- \; g^{\mu\nu} \; + \; \frac{q^{\mu}q^{\nu}}{q^2} \;\; = \;\; 
\sum_{T=T1}^{T2} \; ( \varepsilon^{\ast}_{T} )^{\mu} \; ( \varepsilon_{T} )^{\nu}  \; - \; ( \varepsilon^{\ast}_{L} )^{\mu} \; ( \varepsilon_{L} )^{\nu} \; .
\label{ap3}
\end{equation}
The four polarization vectors of the virtual photon satisfy also
the orthonormality relation:
\begin{equation}
( \varepsilon^{\ast}_{m} )_{\mu} \; ( \varepsilon_{n} )^{\mu}  \;\; = \;\;
\zeta_m \delta_{mn} \; , \;\;\;\;\;\; \mbox{where}
\;\;\; \zeta_L = 1 \; , \;\;\; 
\zeta_S = \zeta_T = -1 \;. 
\label{ap4}
\end{equation}

\noindent
{\large\bf {A.2 Linear polarization}}

The linear-polarization states are represented by  the real polarization vectors.
The  longitudinal-polarization vector 
($\varepsilon^{2}_{L} = 1$) for the electron--proton
collisions depends on the four-momenta of the virtual photon
($q^{\mu}$) and proton ($p_p^{\mu}$) or quark ($p_q^{\mu} = x p_p^{\mu}$), 
(see also \cite{Budnev:1975}):
\begin{equation}
\varepsilon^{\mu}_{L} \;\; = \;\; \frac{(p_pq) q^{\mu} - q^{2} p_p^{\mu}}{
\sqrt{-q^{2}} \; (p_pq)} \; .
\label{43}
\end{equation}
For the semi-inclusive processes we need to construct  
the two transverse-polarization vectors ($\varepsilon^{2}_{T} = -1$).
In  order to construct them we have to introduce the third four-momentum;
for the Compton process we  use the momentum of the final photon
($p^{\mu}$) and  obtain:
\begin{equation}
\varepsilon^{\mu}_{1} \;\; = \;\; \frac{- 2}{\sqrt{stu}} \;
[ \; (p_pp)q^{\mu} -(p_pq)p^{\mu} +
\frac{(qp)(p_pq) - q^{2}(p_pp)}{(p_pq)} \; p_p^{\mu} \; ]
\label{44}
\end{equation}
and
\begin{equation}
\varepsilon^{\mu}_{2} \;\; = \;\; \frac{2}{\sqrt{stu}} \;
\epsilon^{\mu\nu\alpha\beta} \; q_{\nu} (p_p)_{\alpha} p_{\beta},
\label{45}
\end{equation}
\vspace{3 mm}
where $stu = 4 [2(p_pq)(p_pp)(qp) - q^{2}(p_pp)^{2}]$ .\\
This transverse-polarization vectors satisfy in addition the following equations: 
$$\varepsilon_{1} p_p = 0,\;\;\;\;
\varepsilon_{1} p = 0 \;\;\;\; {\rm and} \;\;\;\; \varepsilon_{2} p_p = 0.$$

\noindent
{\large\bf A.3 Circular polarization}

Using the above forms of the linear-polarization vectors 
(eqs. (\ref{43}), (\ref{44}),  (\ref{45})),
one can define the circular-polarization
vectors of the $\gamma^{\ast}$, namely
the longitudinal-polarization 
vector  denoted by $\varepsilon^{\mu}_{0}$ and the transverse ones
denoted by $\varepsilon^{\mu}_{+}$ and $\varepsilon^{\mu}_{-}$.
These  vectors correspond to the helicities
of the intermediate photon equal to $\lambda = 0, +,$ and $ -$, respectively.
We get
\begin{equation}
\varepsilon^{\mu}_{0} \;\; = \;\; i \; \varepsilon^{\mu}_{L},
\label{49}
\end{equation}
\begin{equation}
\varepsilon^{\mu}_{+} \;\; = \;\;  \frac{1}{\sqrt{2}} \;
( \;\varepsilon^{\mu}_{1} \; + \; i \varepsilon^{\mu}_{2} \; ), 
\label{410}
\end{equation}
and
\begin{equation}
\varepsilon^{\mu}_{-} \;\; = \;\; \frac{1}{\sqrt{2}} \;
( \; \varepsilon^{\mu}_{1} \; - \; i \varepsilon^{\mu}_{2} \; ). 
\label{411}
\end{equation}

\noindent 
{\large\bf A.4 Factorization formula for the Compton process}

The general factorization formula for the electron--quark
scattering with the photon in the final state has the following form
for massless quarks:
\begin{equation}
d\sigma^{e q \rightarrow e q \gamma } \; \sim \;
e^6Q_q^4 \; {\rm Re} \; [ L_{e \; \alpha\beta} \frac{g^{\alpha\mu}g^{\nu\beta}}{Q^4} T_{\mu\nu} ]
\; = \; \frac{e^6Q_q^4}{Q^4} \; {\rm Re} \; [ \sum_{i,j} \;  \zeta_{ij}  \;
\underbrace{ (L_{e \; \alpha\beta} \varepsilon_{i}^{\ast \alpha} 
\varepsilon_{j}^{\beta}) }_{ \equiv \; L_{ij} } \;
\underbrace{ (\varepsilon_{j}^{\ast \nu} \varepsilon_{i}^{\mu}
T_{\mu\nu}) }_{ \equiv \; T_{ji} } ]\; ,
\label{eq:mkj:ap1}
\end{equation}
where
$i, j$ denote the polarization states of the virtual photon
($i,j = L, T1, T2$ for the linear polarization and  $i,j = 0, +, -$
for the circular polarization),
$\zeta_{LT}=\zeta_{TL}=-1$, while in the remaining cases $\zeta_{ij}=1$.
The tensor  $L_{e \; \mu\nu} (q, k)$\footnote{
The leptonic tensor $L_{e \; \mu\nu} (q, k)$ 
is given by  eq. (\ref{ll}).}
 is related to the emission of
the virtual photon by the electron, while the tensor
   $T_{\mu\nu} (q, p_q, p)$ describes  the absorption of 
the virtual photon by the quark (from proton), with the subsequent 
production of a prompt photon.

Two of the contributions to the cross section, 
$d \sigma_T$ (\ref{T}) and $d \tau_{TT}$ (\ref{TT}),
depend on the choice of the  basis of the polarization states of 
$\gamma^{\ast}$.
But the propagator decomposition method assures that
their sum: $d \hat{\sigma}_T = d \sigma_T + d \tau_{TT}$, as well as 
the individual terms 
$d \sigma_L$ (\ref{L}) and  $d \tau_{LT}$ (\ref{LT})
are the same in  any basis.
This fact can also be checked
by using the relations 
between the linear and circular 
transverse-polarization vectors  (eqs. (\ref{410}), (\ref{411})).\\

The calculations presented in this paper
rely on  the lowest order 
subprocess $\gamma^{\ast} q \; \rightarrow \; \gamma q$
(the Born approximation).
The corresponding partonic tensor has the following form\footnote{
The tensor for the process:
$e \gamma^{\ast} \rightarrow e \gamma$ at the Born
level with   massive fermions
was calculated in  \cite{Kuraev:1987cg}. The  hadronic tensor $T^{\mu\nu}$
obtained by us
is in agreement with this tensor for  the massless fermions.}:
\begin{eqnarray*}
T^{\mu\nu}(q, p_q, p) \;\; = \;\; 
\frac{2}{su} \; \{ \; g^{\mu\nu} ( u^{2} + s^{2} + 2q^{2}t )
\; + \;
2u[2q^{\mu}q^{\nu} + q^{\mu}(p_q^{\nu}-p^{\nu}) + (p_q^{\mu}-p^{\mu})q^{\nu}] \; +\;
\end{eqnarray*}
\begin{equation}
\; - \; 2s ( p_q^{\mu}q^{\nu} + q^{\mu}p_q^{\nu} )
\; +\; 2q^{2}[4p_q^{\mu}p_q^{\nu} + 2p^{\mu}p^{\nu} +
q^{\mu}(2p_q^{\nu}-p^{\nu}) + (2p_q^{\mu}-p^{\mu})q^{\nu}
- 2p_q^{\mu}p^{\nu} - 2p^{\mu}p_q^{\nu}] \; \} \; ,
\label{34}
\end{equation}
\vspace{5 mm}
where $s$, $t$, $u$ are the Mandelstam variables for the process $\gamma^{\ast} q \; \rightarrow \; \gamma q$:
\begin{equation}
s \; = \; (q+p_q)^{2} \; , \;\;\;\;\;\;
t \; = \; (q-p_{\gamma})^{2} \; , \;\;\;\;\;\;
u \; = \; (p_q-p_{\gamma})^{2} \; .
\label{35}
\end{equation}

\noindent
{\Large\bf Appendix B.}
\noindent
{\Large\bf {        The Breit frame}}

\noindent
{\large\bf B.1 Kinematical variables}

The special reference frame, called
the Breit frame (see also \cite{Kopp:1978vx}),
is defined by choosing the momenta of the exchanged photon  and 
the electrons in the following form:
\begin{equation}
q^{\mu} \;\; = \;\; \sqrt{Q^{2}} \; (0, \; 0, \; 0, \; 1) \;, \;\;\;\;\;\;\;\;
- q^2 \;\; = \;\; Q^2 \; ;
\label{eq:mkj:ap2}
\end{equation}
\begin{equation}
k^{\mu} \;\; = \;\; \frac{1}{2} \sqrt{Q^{2}} \; 
(\cosh \psi, \; \sinh \psi \cos \phi_e, \; \sinh \psi \sin \phi_e, \; 1) \;,
\;\;\;\;\;\;\;\; k^2 \;\; = \;\; 0 \; ;
\label{eq:mkj:ap3}
\end{equation}
\begin{equation}
k^{\prime \mu} \;\; = \;\; \frac{1}{2} \sqrt{Q^{2}} \; 
(\cosh \psi, \; \sinh \psi \cos \phi_e, \; 
\sinh \psi \sin \phi_e, \; -1) \; ,
\;\;\;\;\;\;\;\; k^{\prime \; 2} \;\; = \;\; 0 \; .
\label{eq:mkj:ap4}
\end{equation}
$\phi_e$ is the azimuthal angle of the scattered electron.
The hyperbolic functions of the angle $\psi$ are related to the variables
$y \; = \; qp_p/kp_p$ as follows:
\begin{equation}
\cosh \psi  \;\; = \;\; \frac{1}{y} \; (2 - y) \; , \;\;\;\;\;\;
\sinh \psi \;\; = \;\; \frac{2}{y} \sqrt{1-y} \; .
\label{eq:mkj:ap7}
\end{equation}
The momenta of the initial proton ($p_p^{\mu}$),
the initial quark ($p^{\mu}_q$) and the final photon ($p^{\mu}$)
are given by:
\begin{equation}
p_p^{\mu} \;\; = \;\; (E_p, \; 0, \; 0, \; -E_p) \; , \;\;\; \;\;\;
p_q^{\mu} \;\; = \;\; x \; p_p^{\mu}  
\label{eq:mkj:ap8}
\end{equation}
and
\begin{equation}
p^{\mu} \;\; = \;\; p_{T} \; (\frac{1}{\sin  \theta_{\gamma}}, \; 
\cos \phi_{\gamma}, \; \sin \phi_{\gamma}, \; 
\frac{\cos \theta_{\gamma}}{\sin \theta_{\gamma}} ) \; .
\label{eq:mkj:ap9}
\end{equation}
where $E_p$  is the energy of the initial proton\footnote{This 
is limited by the energy of the initial electron 
and $S_{ep}$ in the $ep$ centre-of-mass frame:
\begin{equation}
S_{ep} \;\; = \;\; ( \; k \; + \; p_p \;)^2 \;.
\end{equation}},
$x$  the Bjorken scaling variable,
$p_{T}$  the transverse momentum of the prompt photon (perpendicular to the
momenta of the initial electron and proton),
$\phi_{\gamma}$  the azimuthal angle of the $\gamma$
and $\theta_{\gamma}$ - the polar angle 
between the direction of the initial electron
and the direction of the final photon.

\noindent
{\large\bf B.2 Polarization vectors of the $\gamma^{\ast}$
in the Breit frame}

The longitudinal polarization vector in the  Breit frame has
a very simple form:
\begin{equation}
\varepsilon^{\mu}_{L} \;\; = \;\; -i \; \varepsilon^{\mu}_{0} \;\; = \;\; 
( \; 1, \; 0, \; 0, \; 0 \; ) \; . 
\label{331}
\end{equation}
The transverse-polarization vectors also simplified
in this frame, but they still depend on a choice of the basis.
For the linear polarization we obtain:
\begin{equation}
\varepsilon^{\mu}_{1} \;\; = \;\; 
( \; 0, \;  \cos \phi_{\gamma}, \;  \sin \phi_{\gamma}, \; 0 \; ) \; ,
\label{332} 
\end{equation}
\begin{equation}
\varepsilon^{\mu}_{2} \;\; = \;\; 
( \; 0, \; - \; \sin \phi_{\gamma}, \;\;\; \cos \phi_{\gamma}, \; 0 \; ) \; .
\label{333} 
\end{equation}
while for the circular polarization we have:
\begin{equation}
\varepsilon^{\mu}_{+} \;\; = \;\; \frac{1}{\sqrt{2}} \;
( \; 0,\;\;\; \cos \phi_{\gamma} \;-\; i \sin \phi_{\gamma}, \;
\;\; \sin \phi_{\gamma} \;+\; i \cos \phi_{\gamma},\;0 \; ) \; ,
\label{334}
\end{equation}
\begin{equation}
\varepsilon^{\mu}_{-} \;\; = \;\; \frac{1}{\sqrt{2}} \;
( \; 0,\; \cos \phi_{\gamma} \;+\; i \sin \phi_{\gamma}, \;
\; \sin \phi_{\gamma} \;-\; i \cos \phi_{\gamma},\;0 \; ) \; .
\label{335}
\end{equation}

\noindent
{\large\bf B.3 Factorization formula}
\label{sec:mkj:623}

From the  momenta and polarization  vectors defined above
one can obtain the explicit form of the coefficients $L_{ij}$ and
$T_{ij}$ (defined in eq. (\ref{eq:mkj:ap1})). They can be treated as the elements of the matrices $L$
and $T$, respectively (the ordering of the rows and columns is 
$T1, L, T2$ for the linear-polarization
 basis and  $+, 0, -$ for the circular-polarization basis).
These matrices calculated in the Born
approximation (in the Breit frame) 
depend on the base considered:
\begin{itemize}
\item Linear polarization
\begin{equation}
\mathbf{L} =  \frac{Q^2}{2}
\left( \begin{array}{ccc}
2 \sinh^{2}\psi \cos^{2}\phi + 2                                & 
- \sinh2\psi  \cos\phi  &
\sinh^{2}\psi \sin2\phi              \\
- \sinh2\psi  \cos\phi   &
2 \sinh ^{2}\psi                                  & 
- \sinh2\psi  \sin\phi \\
\sinh^{2}\psi \sin2\phi                 & 
- \sinh2\psi  \sin\phi   & 
2 \sinh^{2}\psi \sin^{2}\phi + 2 
\end{array} \right)
\label{eq:mkj:ap13}
\end{equation}
and
\begin{equation}
\mathbf{T} =  2 \frac{Q^{2}}{su}
\left( \begin{array}{ccc}
- U - 4 p_{T}^{2}                        & 
- 4 p_{T} \hat{E}                                  &
0                                                     \\
- 4 p_{T} \hat{E}                                   &
U - 4 (\hat{E}^{2} + E_q^{2}) & 
0                               \\
0                                                        & 
0                                & 
- U
\end{array} \right) \; ,
\label{eq:mkj:ap14l}
\end{equation}
\end{itemize}
\begin{itemize}
\item Circular polarization
\begin{equation}
\mathbf{L} =  \frac{Q^2}{2}
\left( \begin{array}{ccc}
\cosh ^{2} \psi + 1                                & 
- \frac{i}{\sqrt{2}} \sinh 2 \psi \; e^{-i \phi}  &
\sinh ^{2} \psi \; e^{-2 i \phi}              \\
\frac{i}{\sqrt{2}} \sinh 2 \psi \; e^{i \phi}   &
2 \sinh ^{2} \psi                                  & 
\frac{i}{\sqrt{2}} \sinh 2 \psi \; e^{-i \phi} \\
\sinh ^{2} \psi \; e^{2 i \phi}               & 
-\frac{i}{\sqrt{2}} \sinh 2 \psi \; e^{i \phi}  & 
\cosh ^{2} \psi + 1 
\end{array} \right)
\label{eq:mkj:ap13}
\end{equation}
and
\begin{equation}
\mathbf{T} =  2 \frac{Q^{2}}{su}
\left( \begin{array}{ccc}
- U  - 2 p_{T}^{2}                     & 
- i 2\sqrt{2} p_{T} \hat{E}                                 &
-2p_{T}^{2}                                                       \\
i 2\sqrt{2} p_{T} \hat{E}                                 &
U - 4 (\hat{E}^{2} + E_q^{2}) & 
i 2\sqrt{2} p_{T} \hat{E}                               \\
-2p_{T}^{2}                                                        & 
- i 2\sqrt{2} p_{T} \hat{E}                                & 
- U - 2 p_{T}^{2} 
\end{array} \right) \; ,
\label{eq:mkj:ap14c}
\end{equation}
\end{itemize}
where $\hat{E} = E_q - E_{\gamma}$ ($E_q = x E_p$, $E_{\gamma} = \frac{p_T}{\sin\theta\gamma}$), $U =  \frac{u^{2}+s^{2}}{Q^{2}} - 2t$
and $\phi \; = \; \phi_e - \phi_{\gamma}$.\\

\noindent
{\large\bf B.4 The azimuthal-angle distribution}

In  both bases
the azimuthal-angle distribution for the Compton process (in the Born approximation)
depends only on $\cos \phi$ and
$\cos 2\phi$, namely:
\begin{equation}
\frac{d\sigma^{e q \rightarrow e q \gamma }}{d\phi} 
\; = \; 
\sigma_{0} + \sigma_{1} \cos \phi + \sigma_{2} \cos 2\phi \; .
\label{eq:mkj:ap16}
\end{equation}
It is only for the circular-polarization  basis 
that the coefficients $\sigma_{i},( i=0,1,2)$ in formula (\ref{eq:mkj:ap16})
are strictly related to the corresponding four contributions
$d \sigma_T$, $d \sigma_L$, $d \tau_{LT}$ and $d \tau_{TT}$, calculated in the Breit frame: 
\begin{equation}
\sigma_0 \; = \; 
\frac{d \sigma_T}{d \phi} + \frac{d \sigma_L}{d \phi} 
\; \sim \; - 4 \frac{1}{su} \;
[\; (\cosh ^{2} \psi + 1)\;p_{T}^{2} \;+\; 2 \sinh ^{2} \psi \; 
[\hat{E}^{2} + E_q^{2}] \; + \; U   \;] \;,
\label{eq:mkj:ap18}
\end{equation}
\begin{equation}
\sigma_1 \; \cos \phi \; = \;
\frac{d \tau_{LT}}{d \phi} 
\; \sim \; 
\frac{8}{su} \;
\sinh 2\psi \; p_{T} \; \hat{E} \; \cos \phi \; ,
\label{eq:mkj:ap20}
\end{equation}
\begin{equation}
\sigma_2 \; \cos 2\phi \; = \; 
\frac{d \tau_{TT}}{d \phi} 
\; \sim \;
\frac{-4}{su} \;
\sinh ^{2} \psi \; p_{T}^{2} \; \cos 2\phi \;.
\label{eq:mkj:ap22}
\end{equation}

The longitudinal--transverse interference does not depend on the choice 
of the basis for the polarization vectors and consequently, it 
 is related to the  $\cos \phi$ term also
for the linear-polarization vectors.


\vspace{1cm}
\begin{thebibliography}{99}

\bibitem{Hand:1963bb}
L.~N.~Hand,
Phys.\ Rev.\  {\bf 129} (1963) 1834.

\bibitem{Hand:1961bv}
L.~N.~Hand, Ph.D. thesis (Stanford U., Phys. Dept., 1961)
RX-933.

\bibitem{Gourdin:1961}
M.~Gourdin,
Nuovo\ Cim.\ {\bf 21} (1961) 1094.

\bibitem{Dalitz:1957}
R.H.~Dalitz and D.R.~Yennie,
Phys.\ Rev.\ {\bf 105} (1957) 1598.

\bibitem{Yennie:1957}
D.R.~Yennie, M.M.~Levy and D.G.~Ravenhall,
Rev.\ Mod.\ Phys.\ {\bf 29} (1957) 144.

\bibitem{Budnev:1975}
V.M.~Budnev, I.F.~Ginzburg, G.V.~Meledin and V.G.~Serbo,
Phys.\ Rep.\ {\bf 15C} (1975) 181.

\bibitem{Dobrovolskaya:1991kx}
A.~Dobrovolskaya and V.~Novikov,
Z.\ Phys.\ C {\bf 52} (1991) 427.

\bibitem{Bambade:1993vw}
P.~Bambade, A.~Dobrovolskaya and V.~Novikov,
Phys.\ Lett.\ B {\bf 319} (1993) 348.

\bibitem{Arteaga:1995px}
N.~Arteaga, C.~Carimalo, P.~Kessler, S.~Ong and O.~Panella,
Phys.\ Rev.\ D {\bf 52} (1995) 4920
[hep-ph/9511398].

\bibitem{Abbiendi:1999pv}
G.~Abbiendi { et al.}  [OPAL Collaboration],
Eur.\ Phys.\ J.\ C {\bf 11} (1999) 409
[hep-ex/9902024].

\bibitem{Nisius:2000cv}
R.~Nisius,
Phys.\ Rep.\ {\bf 332} (2000) 165
[hep-ex/9912049].

\bibitem{Krawczyk:2000nh}
M.~Krawczyk,
hep-ph/0012179.

\bibitem{Berger:1984xu}
C.~Berger { et al.}  [PLUTO Collaboration],
Phys.\ Lett.\ B {\bf 142} (1984) 119.

\bibitem{L3:2000}
M.~Acciarri et al., [L3]
Phys.\ Lett.\ B {\bf 483} (2000) 373

\bibitem{Schienbein:2002wj}
I.~Schienbein,
arXiv:hep-ph/0205301.

\bibitem{Friberg:2000zz}
C.~Friberg and T.~Sj\"ostrand,
Eur.\ Phys.\ J.\ C {\bf 13} (2000) 151
[hep-ph/9907245].

\bibitem{Friberg:2000dc}
C.~Friberg,
Nucl.\ Phys.\ Proc.\ Suppl.\ {\bf 82} (2000) 124
[hep-ph/9907299].

\bibitem{Friberg:2000pd}
C.~Friberg,
hep-ph/0005048.

\bibitem{Friberg:2000ra}
C.~Friberg and T.~Sj\"ostrand,
JHEP{\bf 0009} (2000) 010
[hep-ph/0007314].

\bibitem{Friberg:2000nx}
C.~Friberg and T.~Sj\"ostrand,
Phys.\ Lett.\ B {\bf 492} (2000) 123
[hep-ph/0009003].

\bibitem{Friberg:2000uq}
C.~Friberg,
hep-ph/0010264.

\bibitem{Chyla:2000cu}
J.~Chyla and M.~Tasevsky,
Eur.\ Phys.\ J.\ C {\bf 16} (2000) 471
[hep-ph/0003300].

\bibitem{Chyla:2000hp}
J.~Chyla,
Phys.\ Lett.\ B {\bf 488} (2000) 289
[hep-ph/0006232].

\bibitem{Chyla:2000es}
J.~Chyla,
hep-ph/0010055.

\bibitem{Chyla:2001ue}
J.~Chyla and M.~Tasevsky,
Eur.\ Phys.\ J.\ C {\bf 18} (2001) 723
[hep-ph/0010254].

\bibitem{Kessler:1970ef}
P.~Kessler,
Nucl.\ Phys.\ B {\bf 15} (1970) 253.

\bibitem{ujd:1999}
U.~Jezuita-D\c{a}browska,  MSc Thesis (1999)
 Warsaw University

\bibitem{Brown:1971}
R.~Brown and I.~Muzinich,
Phys.\ Rev.\ D {\bf 4} (1971) 1496.

\bibitem{Georgi:1978}
H.~Georgi and H.D.~Politzer,
Phys.\ Rev.\ Lett. {\bf 40} (1978) 3.

\bibitem{Cahn:1978se}
R.~N.~Cahn,
Phys.\ Lett.\ B {\bf 78} (1978) 269.

\bibitem{Kopp:1978vx}
G.~Kopp, R.~Maciejko and P.~M.~Zerwas,
Nucl.\ Phys.\ B {\bf 144} (1978) 123.

\bibitem{Mendez:1978zx}
A.~Mendez,
Nucl.\ Phys.\ B {\bf 145} (1978) 199.

\bibitem{Chay:1991jc}
J.~Chay, S.~D.~Ellis and W.~J.~Stirling,
Phys.\ Lett.\ B {\bf 269} (1991) 175;
Phys.\ Rev.\ D {\bf 45} (1992) 46.

\bibitem{Oganesian:1998jq}
K.~A.~Oganesian, H.~R.~Avakian, N.~Bianchi and P.~Di Nezza,
Eur.\ Phys.\ J.\ C {\bf 5} (1998) 681
[hep-ph/9709342].

\bibitem{Ahmed:1999ix}
M.~Ahmed and T.~Gehrmann,
Phys.\ Lett.\ B {\bf 465} (1999) 297
[hep-ph/9906503].

\bibitem{Lai:2000wy}
H.~L.~Lai { et al.}  [CTEQ Collaboration],
Eur.\ Phys.\ J.\ C {\bf 12} (2000) 375
[hep-ph/9903282].

\bibitem{Brodsky:1972yx}
S.~J.~Brodsky, J.~F.~Gunion and R.~Jaffe,
Phys.\ Rev.\ D {\bf 6} (1972) 2487.

\bibitem{Brodsky:1999sr}
S.~J.~Brodsky, M.~Diehl, P.~Hoyer and S.~Peigne,
Phys.\ Lett.\ B {\bf 449} (1999) 306
[hep-ph/9812277].

\bibitem{Hoyer:2000mb}
P.~Hoyer, M.~Maul and A.~Metz,
Eur.\ Phys.\ J.\ C {\bf 17} (2000) 113
[hep-ph/0003257].

\bibitem{Kramer:1998nb}
G.~Kramer, D.~Michelsen and H.~Spiesberger,
Eur.\ Phys.\ J.\ C {\bf 5} (1998) 293
[hep-ph/9712309].

\bibitem{Breitweg:2000qh}
J.~Breitweg { et al.}  [ZEUS Collaboration],
Phys.\ Lett.\ B {\bf 481} (2000) 199
[hep-ex/0003017].

\bibitem{Kuraev:1987cg}
E.~A.~Kuraev, N.~P.~Merenkov and V.~S.~Fadin,
Yad.\ Fiz.\  {\bf 45} (1987) 782
[Sov.\ J.\ Nucl.\ Phys.\  {\bf 45} (1987) 486].

\end{thebibliography}
\end{document}